\documentclass[12pt]{article}

\usepackage[dvips,letterpaper]{geometry}
\usepackage{graphicx,psfrag,epsf}
\usepackage{times}
\usepackage{fullpage}
\usepackage{setspace}
\usepackage{amsmath,amssymb,amsfonts}
\usepackage{bm}
\usepackage{graphicx}
\usepackage{epsfig}
\usepackage{subfigure}
\usepackage{url}
\usepackage{verbatim}
\usepackage{natbib}
\usepackage{color}
\usepackage{booktabs,multirow}
\usepackage{colortbl}
\usepackage{graphicx,psfrag,epsf}
\usepackage{algorithm,algpseudocode}
  \usepackage{hhline}

\title{\bf {\Large The negative binomial beta prime regression model with cure rate}}

\author{
\normalsize
\textbf{Jeremias Le\~ao}$^{1}$,\,\textbf{Marcelo Bourguignon}$^{2}$,\,\textbf{Manoel Santos-Neto}$^{3}$\,\textbf{and}\,
\textbf{Helton Saulo}$^{4}$\\[-0.15cm]
{\small $^{1}$Department of Statistics, Universidade Federal do Amazonas, Brazil}\\[-0.15cm]
{\small $^{2}$Department of Statistics, Universidade Federal do Rio Grande do Norte, Brazil}\\[-0.15cm]
{\small $^{3}$Department of Statistics, Universidade Federal de Campina Grande, Brazil}\\[-0.15cm]
{\small $^{4}$Department of Statistics, Universidade de Bras\'ilia, Brazil}\\
}

\date{}

\begin{document}

\maketitle
\begin{abstract}
\noindent
This paper introduces a cure rate survival model by assuming
that the time to the event of interest follows a beta 
prime distribution and that the number of competing causes of the event of interest follows a negative binomial distribution. This model provides a novel alternative to the existing cure rate regression models due to its flexibility, as the beta prime model can exhibit greater levels of skewness and kurtosis than those of the gamma and inverse Gaussian distributions. Moreover, the hazard rate of this model can have an upside-down bathtub or an increasing shape. We approach both parameter estimation and local influence based on likelihood methods. In special, three perturbation schemes are 
considered for local influence. Numerical evaluation of the proposed model is performed by Monte Carlo simulations. In order to illustrate the potential for practice of our model we
apply it to a real data set.
\paragraph{Keywords}
Beta prime distribution; Likelihood methods; Local influence; Long-term survival model; Medical data;
Negative binomial distribution.

\end{abstract}

\section{Introduction}\label{sec1}
\noindent

In medical and epidemiological studies, often interest focuses on
studying the effect of concomitant information on the time to event such as
death or recurrence of a disease. When the primary interest
is to estimate the covariate effect, the Cox proportional hazards model
is commonly used in the analysis of survival time data; see \cite{c:72}.
With the development of medical and health sciences, the datasets collected from
clinical studies pose some new challenges to statisticians. New statistical
models which can incorporate these changes should be investigated. The most
prevalent change noted in many clinical studies is that, more patients
respond favorably to a treatment or, were not susceptible to the event
of interest in the study, so they are considered cured or have prolonged
disease-free survival. This proportion of patients is called the cure fraction.
Incorporating the cure fraction in survival models leads to cure rate
models or long-term survival models. These models have been widely developed
in the biostatistics literature. Historically, one of the most famous
cure rate models is the mixture cure model introduced by \cite{bg:52}.
This model has been extensively discussed by several authors, including \cite{f:82},
\cite{mazhou:96}, \cite{ei:97}, \cite{sg:98} and
\cite{ctf:13}. Later, \cite{yt:96} and \cite{cis:99}
proposed the promotion time cure model or bounded cumulative hazard model in cancer relapse settings, assuming that a latent biological process of propagation of latent carcinogenic tumor cells is generating the observed failure (relapse). Recently, \cite{cbcs:07} generalized this framework to a flexible class of cure models under latent activation schemes, \cite{canchoetal2011} proposed a flexible cure rate model, which encompasses as special cases the mixture model \citep{bg:52}, the promotion time cure model \citep{cis:99}, and the cure rate proportional odds model proposed by \cite{gsb:11}. The statistical literature for modeling lifetime data in the presence of a cure fraction and latent competing causes is vast and growing rapidly. Interested readers can refer to  \cite{yt:96}, \cite{tiy:03}, \cite{yi:05}, \cite{cbcs:07}, \cite{cetal:09}, \cite{rgetal:09}, \cite{canchoetal2012}, \cite{brl:12}, among others.

In this context, our main objective is to introduce the negative binomial beta prime
(NBBP) cure rate model, conceived inside a latent competing causes
scenario with cure fraction, where there is no information about which
cause was responsible for the individual death or tumor recurrence,
but only the minimum lifetime value among all risks is observed and
a part of the population is not susceptible to the event of interest.
The beta prime model has properties that its competitor distributions of
the exponential family do not have. For example, the hazard rate function of beta prime distribution can have an upside-down bathtub or increasing shape depending on the parameter values. Most classical two-parameter distributions such as Weibull and gamma distributions have monotone hazard rate
functions. The skewness and kurtosis of the beta prime distribution can be much larger than those of the gamma and inverse
Gaussian distributions, which may
be more appropriate in certain practical situations.

The rest of this paper is organized as follows.
In Section~\ref{sec2}, we introduce
the proposed long-term survival model
and discuss some of its properties as well as
some special models. The estimation method for the model parameters
is discussed in Section~\ref{sec_inf}.
In Section~\ref{sec:num_app}, we illustrate
the proposed model through simulation and
an application to medical real-world data set.
In Section~\ref{sec_conc}, we mention some concluding remarks.

\section{The negative binomial beta prime regression model with cure rate}\label{sec2}

In this section we will formulate the NBBP regression model with cure rate. The ingredients of the proposed model are: (i) the unified long-term survival model used formulated by
\cite{rgetal:09}; (ii) the BP distribution for modeling the time to the event of interest; and (iii) the NB distribution for modeling the number of competing causes of the event of interest.

Let $N$ denote the number of competing causes related to the occurrence
of an event of interest, for an individual in the
population. Conditional on $N$, we assume that the $Z_j$'s are independent and identically distributed random variables representing
the promotion times of the competing causes, for $j=1, \ldots, n$. Moreover, we assume that $N$ is independent of $Z_{1},\ldots, Z_{n}$ and the
observable time-to-event is defined as $T=\min\{Z_{1}, \ldots, Z_{N}\}$ for $N\geq 1$,
and $T=\infty$ if $N=0$, which leads to a cured fraction denoted by $p_{0}$; see \cite{rgetal:09}. Under this setup, the long-term survival function (SF) of the random variable $T$ is given by
\begin{eqnarray}\label{eqq.4.1}
S_{p}(t|\cdot)&=& \textrm{P}(T\geq t) = \textrm{P}(N=0)+\displaystyle\sum_{n=1}^{\infty}\textrm{P}(Z_{1}>t,\ldots,Z_{N}>t|N=n)\textrm{P}(N=n)\nonumber\\
&=&\displaystyle\sum_{n=0}^{\infty}\textrm{P}(N=n) [S_{T}(t|\cdot)]^{n}=A_{N}(S_{T}(t|\cdot)),\quad t > 0,
\end{eqnarray}
where $S_{T}(\cdot|\cdot)$ denotes the SF of the unobserved lifetimes and $A_{N}(\cdot)$
is the probability generating function of the random variable $N$, which converges when $S_{T}(t|\cdot)\in [0,1]$.
Various results can be obtained for each choice of $A_{N}(\cdot)$ and $S_{T}(\cdot|\cdot)$
considered in \eqref{eqq.4.1}.

A random variable $X$ follows the beta prime (BP) distribution with
shape parameters $\alpha > 0$ and $\beta > 0$,
if its cumulative distribution function (CDF) and
probability density function (PDF) are given by
\begin{align}\label{eq:cdf}
 F_X(x|\alpha,\beta) = I_{x/(1+x)}(\alpha,\beta),
\end{align}
and
\begin{align*}\label{eq:pdf}
 f_X(x|\alpha,\beta) = \frac{x^{\alpha-1}(1+x)^{-(\alpha+\beta)}}{B(\alpha,\beta)},
\end{align*}
where $I_{y}(\alpha,\beta)=B_{y}(\alpha,\beta)/B(\alpha,\beta)$
is the incomplete beta function ratio, $B_{y}(\alpha,\beta)=\int^{y}_{0}u^{\alpha-1}(1-u){\beta-1}{\rm d}u$
is the incomplete function, $B(\alpha, \beta) = \Gamma(\alpha)\Gamma(\beta)/\Gamma(\alpha +\beta)$
is the beta function and $\Gamma(\alpha) =\int^{\infty}_{0} u^{\alpha-1} \exp (-u) {\rm d}u$
is the gamma function. The BP is related with several models.
The interested reader in BP model is referred to \cite{bsm:18,McDBt:90,McD:87,McD:84} and \cite{jkb:95}.
These works present reviews and generalizations of BP model.
In this context, to introduced the negative binomial beta prime (NBBP)
model, we are considering the parameterization used in \cite{bsm:18}, where the PDF
of the BP distribution is given by
\begin{equation}\label{eq:pdf1}
 f_{\rm BP}(t|\mu,\phi) = \frac{t^{\mu(\phi+1)-1}(1+t)^{-[\mu(\phi+1)+\phi+2]}}{B(\mu(\phi+1),\phi+2))},
\end{equation}
where $\alpha=\mu(\phi+1)$ and $\beta=\phi+2$. In this case, the BP distribution is indexed by
in terms of the mean ($\mu$) and precision parameters ($\phi$).

Consider that the number of competing causes $N$ follows a NB distribution (particular cases are the Poisson, binomial , Bernoulli and geometric distributions)
with parameters $\alpha$ and $\theta$, for $\theta>0$ and $\alpha\,\theta>-1$, and that the time to the event of interest is BP distributed with parameters
$\mu$ and $\phi$ as in \eqref{eq:pdf1}. Then, the long-term SF of cured patients is given by
\begin{equation}\label{eqq.4.10}
 S_{p}(t|{\bm \xi}) = \left[1 + \alpha\,\theta F_{\rm BP}(t|\mu,\phi)\right]^{-1/\alpha}, \quad t > 0,
\end{equation}
where ${\bm \xi}=(\alpha,\theta, \mu, \phi)^{\top}$. The corresponding
PDF and HR obtained from \eqref{eqq.4.10} are respectively expressed as
\begin{eqnarray*}
 f_{p}(t| {\bm \xi}) &=& \frac{\theta t^{\mu(\phi+1)-1}(1+t)^{-[\mu(\phi+1)+\phi+2]}}{B(\mu(\phi+1),\phi+2)}\left[1+\alpha\,\theta F_{\rm BP}(t|\mu,\phi)\right]^{-1/\alpha-1} ,\label{eqq.4.11}\\
 \nonumber \\
 h_{p}(t| {\bm \xi}) &=&\frac{\theta t^{\mu(\phi+1)-1}(1+t)^{-[\mu(\phi+1)+\phi+2]}}{B(\mu(\phi+1),\phi+2)}\left[1+\alpha\,\theta F_{\rm BP}(t|\mu,\phi)\right]^{-1},\quad t > 0.\label{eqq.4.12}
\end{eqnarray*}

The SF for the non-cured population (or NBBP SF), denoted by $S_{\rm NBBP}$,
is given by
\begin{equation}\label{eqq.4.11}
 S_{{\rm NBBP}}(t| {\bm \xi}) = \frac{\left[1 + \alpha\,\theta F_{\rm BP}(t|\mu,\phi)\right]^{-1/\alpha}-(1+\alpha\,\theta)^{-1/\alpha}}{1-(1+\alpha\,\theta)^{-1/\alpha}}, \quad t > 0.
\end{equation}

From \eqref{eqq.4.11} we have $\lim\limits_{t \to 0}S_{{\rm NBBP}}(t|{\bm \xi})=1$ and $\lim\limits_{t \to +\infty}S_{{\rm NBBP}}(t|{\bm \xi})=0$,
so $S_{{\rm NBBP}}$ is a proper SF. The PDF of non-cured population (or NBBP PDF),
denoted by $f_{{\rm NBBP}}(t|{\bm \xi})$ is given by
\begin{equation}\label{eqq.4.12}
 f_{{\rm NBBP}}(t| {\bm \xi}) = \frac{\theta f_{{\rm BP}}(t|\mu,\phi)\left[1+\alpha\,\theta F_{\textrm{BP}}(t|\mu,\phi)\right]^{-(1+1/\alpha)}}{1-(1+\alpha\,\theta)^{-1/\alpha}}, \quad t > 0.
\end{equation}

\section{Estimation and Diagnostics}\label{sec_inf}
\noindent

Here, we use the maximum likelihood (ML) method to estimate the model parameters. We assume that the time to event is not completely observed and is subject to
right censoring. We observe $t_{i}=\min\{y_{i},c_{i}\}$ and $\delta_{i}=\mathbb{I}(y_{i}\leq c_{i})$,
where $c_{i}$ is the censoring time, and $\delta_{i}=1$ if $y_{i}$ it is time to the event and $\delta_{i}=0$ if it is right censored, for $i=1,\ldots,n$. Then,
from $n$ pairs of times and censoring indicators, $(y_{1} , \delta_{1} ), \ldots , (y_{1} , \delta_{n} )$ say, the corresponding likelihood function,
under uninformative censoring, is given by
\begin{align}\label{sec_inf:eq1}
L({\bm \vartheta}|{\bm y}) = \prod^{n}_{i=1}[f_{\rm pop}(y_{i}|{\bm \vartheta})]^{\delta_{i}}[S_{\rm pop}(y_{i}|{\bm \vartheta})]^{1-\delta_{i}},
\end{align}
where ${\bm \vartheta}=({\bm \xi},{\bm \beta})^{\top}$,
\begin{equation}\label{sec_inf:eq2}
S_{\rm pop}(y_{i}|{\bm \xi},{\bm \beta}) = \left\{ \begin{array}{ll}
         [1 + (p^{-\alpha}_{0i}-1)F_{\rm BP}(y_{i}|{\bm \xi})]^{-1/\alpha},& \mbox{if $\alpha \neq 0$};\\
         &\\
        p^{F_{\rm BP}(y_{i}|{\bm \xi})}_{0i}, & \mbox{if $\alpha = 0$},\end{array} \right.
\end{equation}
and
\begin{equation}\label{sec_inf:eq3}
f_{\rm pop}(y_{i}|{\bm \xi},{\bm \beta}) = \left\{ \begin{array}{ll}
         [1 + (p^{-\alpha}_{0i}-1)F_{\rm BP}(y_{i}|{\bm \xi})]^{-1/\alpha-1}\left(\frac{p^{-\alpha}_{0i}-1}{\alpha}\right)f_{\rm BP}(y_{i}|{\bm \xi}),& \mbox{if $\alpha \neq 0$};\\
         &\\
        -\log\left(p_{0i}\right)p^{F_{\rm BP}(y_{i}|{\bm \xi})}_{0i}f_{\rm BP}(y_{i}|{\bm \xi}), & \mbox{if $\alpha = 0$,}\end{array} \right.
\end{equation}
with $\log[p_{0i}/(1-p_{0i})]={\bm x}^{\top}_{i}{\bm \beta}$, that is,
$p_{0i} = \frac{\exp({\bm x}^{\top}_{i}{\bm \beta})}{1 + \exp({\bm x}^{\top}_{i}{\bm \beta})}$, $i=1,\ldots,n$, where ${\bm \beta}$ is the vector
of regression coefficients. Therefore, covariates are used to estimate the cured fraction ($p_{0}$).

From \eqref{sec_inf:eq2}-\eqref{sec_inf:eq3},
the likelihood function in \eqref{sec_inf:eq1}
is expressed as

\begin{align}\label{sec_inf:eq4}
L({\bm \vartheta}|{\bm y}) =
\left\{ \begin{array}{ll}
       \prod^{n}_{i=1}  \left[\left(\frac{p^{-\alpha}_{0i}-1}{\alpha}\right)f_{\rm BP}(y_{i}|{\bm \xi})\right]^{\delta_{i}}[1 + (p^{-\alpha}_{0i}-1)F_{\rm BP}(y_{i}|{\bm \xi})]^{-\delta_{i}-1/\alpha},& \mbox{if $\alpha \neq 0$};\\
       &\\
       \prod^{n}_{i=1} [-\log\left(p_{0i}\right)f_{\rm BP}(y_{i}|{\bm \xi})^{\delta_{i}}]p^{F_{\rm BP}(y_{i}|{\bm \xi})}_{0i}, & \mbox{if $\alpha = 0$}.\end{array} \right.
\end{align}

The ML estimators will be obtained using numerical methods, since equating the first-order log-likelihood derivatives to zero leads us to a complicated system of nonlinear equations. It can be easily performed by using standard non-linear maximization procedures found in most of statistical and data analysis packages.


In order to assess the sensitivity of the ML estimators to atypical cases, we perform local influence analysis which is based on the curvature of the log-likelihood function. Recall that ${\bm \vartheta}=({\bm \xi}, \bm{\beta})^{\top}$ and let the vector of perturbations $\bm{\omega}$
be a subset of $\Omega \in\mathbb{R}^m$, whereas $\bm{\omega}_0$ is a non-perturbation vector such that $\ell(\bm{\vartheta}|\bm{\omega}_0) = \ell(\bm{\vartheta}) = \log(L({\bm \vartheta})) $, for all $\bm{\vartheta}$. Then, the likelihood distance (LD) is given by $\textrm{LD}(\bm{\vartheta}) = 2({\ell}(\widehat{\bm{\vartheta}}) - {\ell}(\widehat{\bm{\vartheta}}_{\bm \omega}))$, where $\widehat{\bm{\vartheta}}_{\bm \omega}$ is the ML estimate of $\bm{\vartheta}$ under the perturbed model, and the normal curvature for $\widehat{\bm{\vartheta}}$, at the direction vector $\bm{d}$ ($||\bm{d}||=1$), is given by $C_{\bm d}(\widehat{\bm{\vartheta}}) = 2| \bm{d}^\top {\bm\nabla}^\top {\bm \Sigma}(\widehat{\bm \vartheta})^{-1}{\bm\nabla}\,\bm{d}|$,
where ${\bm\nabla}$ is a $(q+3) \times m$ matrix that depends on the perturbation scheme, whose elements
are $\nabla_{ji} = \partial^2{\ell}(\bm \vartheta| \bm{\omega}) / \partial\vartheta_j\partial\omega_i$,
evaluated at $\bm{\vartheta} = \widehat{\bm{\vartheta}}$ and $\bm{\omega} = \bm{\omega}_0$,
for $j=1, \ldots, q+3$ and $i=1,\dots,m$; see \cite{c:86}. Index plot of the eigenvector $\bm{d}_{\max}$ associated with the maximum eigenvalue
of $\bm{B}({\bm\vartheta}) = -{\bm\nabla}^\top\bm \Sigma({\bm \vartheta})^{-1}{\bm\nabla}$,
$C_{\bm d_{\max}}({\bm\vartheta})$ say, evaluated at $\bm{\vartheta} = \widehat{\bm{\vartheta}}$
and $\bm{\omega} = \bm{\omega}_0$, can indicate cases which have a high influence on $\textrm{LD}({\bm\vartheta})$. Moreover, the vector $\bm{d}_i = \bm e_{in}$ can
be considered to detect local influence, where $\bm{e}_{im}$ denotes an $m \times 1$ vector of zeros with one at the $i$th position. Thus,
the corresponding normal curvature takes the form $C_i(\bm{\vartheta}) = 2 |b_{ii}(\bm{\vartheta})|$,
where $b_{ii}(\bm{\vartheta})$ is the $i$th diagonal element of $\bm{B}({\bm\vartheta})$, for $i=1,\dots,m$, evaluated
at $\bm{\vartheta} = \widehat{\bm{\vartheta}}$ and $\bm{\omega} = \bm{\omega}_0$. Then, the case $i$ is potentially influential if
$C_{i}(\widehat{\bm{\vartheta}}) > \frac{2}{m}\sum^{m}_{i=1}C_i(\widehat{\bm{\vartheta}})$. In this paper, we consider the following
perturbation schemes: case-weight, response and explanatory variable; see \cite{c:87}.

\section{Numerical applications}\label{sec:num_app}
\noindent

This section presents a simulation study
to evaluate the performance of the ML estimators
of the model parameters and an application to a real-world medical data set regarding a study on cutaneous melanoma (a type of malignant cancer), for the evaluation of postoperative
treatment performance by means of a high dose of a certain drug (interferon alfa-2b) for the prevention of recurrence. The patients in this study were added from 1991 to 1995 with a follow-up of 3 years; see \cite{ics:01a}.

\newpage 

\subsection{A simulation study}\label{subsec:num_app}
\noindent

We carry out a Monte Carlo (MC) simulation study to evaluate
the performance of the ML estimators for the proposed model.
The simulation scenario considers the following:
sample sizes $n\in\{200, 400, 600, 800, 1000\}$,
$\mu = \{0.50,1.00\}$, $\sigma=\{1.00,10.00\}$ and 5,000 MC replications. The cured fraction
is $p_{0i}=\exp(b_{0}+b_{1}x_{i})/[1+\exp(b_{0}+b_{1}x_{i})]$.
For the simulations, we consider a binary covariate $x$
with values drawn from a Bernoulli
distribution with parameter 0.5.
We consider $b_{0} = 0.50$ and $b_{1} = -1.00$
such that the cure fraction for the two levels
of $x$ are $p_{00} = 0.62$ and $p_{01} = 0.38$, respectively.
The censoring times were samples from the uniform
distribution, $\textrm{U}(a, b)$, where $a,b>0$
were set in order to control the proportion of censored observations.
In our study, the proportion of censored observations was on the average
obtained for the each sample size; see Table~\ref{tab:1}.
Note that, based on the probability integral transform, the
NBBP model CDF follows a $U(0,1)$ distribution. Then, the NBBP SF is $U(0,1)$ distributed as well. Random number generation from the NBBP model is performed following Algorithm 1. In step \#2 of this algorithm, we use the function \texttt{uniroot} of the \texttt{R} software to get the root of the equation;
see Brent (1973). For each value of the parameter, sample size and censoring proportion, we report the empirical values for the bias and mean squared error (MSE) of the ML estimators in Table~\ref{tab:1}. From this table,
note that, as the sample size increases, the ML estimators become more efficient, as expected.

\begin{algorithm}[H]
\small
\caption{\small Generator of random numbers from the NBBP model.}\label{alg:1}
\begin{algorithmic}[1]
\State
Fix the parameters values;
\State
Obtain a random number of the
covariate ${\bm x}$ from $X \sim \textrm{Bernoulli}(1,1/2)$,
$V \sim U(0,1)$ and $W \sim U(p_{0j},1)$, $j=\{0,1\}$;

\State
Let $X_{i}=j$. If $v_{i}<p_{0j}$, $t=\infty$, otherwise

$$[1 + (p^{-\alpha}_{0i}-1)F_{BP}(y_{i}|{\bm \xi})]^{-1/\alpha}=w_{i},$$
\noindent
and
$$p^{F_{BP}(y_{i}|{\bm \xi})}_{0i}=w_{i},$$
if $\alpha\neq 0$ and $\alpha = 0$, respectively;
\State
Extract the censored time $c_{i}$ from $C \sim U(a, b)$, for $a,b > 0$ fixed;

\State
Compute $t_{i} = \min\{y_{i}, c_{i}\}$;

\State
If $y_{i}<c_{i} $, then $\delta_{i}=1$, otherwise $\delta_{i}=0$;

\State
Repeat Steps 1 to 6 until the required number of data has been generated.
\end{algorithmic}
\end{algorithm}

\begin{table}[!h]
 \small
\renewcommand{\arraystretch}{0.9}
\renewcommand{\tabcolsep}{0.1cm}
\centering

\caption{empirical mean and SD of the ML estimators of
cure fractions for simulated data from the NBBP model.}\label{tab:1}

\begin{tabular}{lcccccccc}
\hline
  \multirow{2}{*}{$n$ (censoring \%)}    & \multicolumn{7}{c}{EM} \\  \cline{2-8}
  & $\widehat{\mu}$& $\widehat{\phi}$& $\widehat{\alpha}$& $\widehat{\beta}_{0}$& $\widehat{\beta}_{1}$& $\widehat{p}_{00}$& $\widehat{p}_{01}$ \\
 \hline
\multicolumn{1}{c}{\textbf{True Values} $\to$} & $\mathbf{0.500}$&$\mathbf{1.000}$&$\mathbf{2.000}$&$\mathbf{0.500}$&$\mathbf{-1.000}$&$\mathbf{0.623}$&$\mathbf{0.377}$\\

200  (52.98\%)  & 0.5441 & 1.3714 & 2.4005 & 0.4410 & $-$1.0482 &  0.6076  &  0.3571 \\
& (0.3381) & (1.2152) & (1.5111) & (0.2334) & (0.2928) & (0.8946) & (0.5843)\\

400  (52.63\%) & 0.5381 & 1.1324 & 2.1477 & 0.4781 & $-$1.0324 & 0.6169   & 0.3598\\
& (0.2103) & (0.7293) & (0.9810) & (0.1598) & (0.2061)  & (0.5929) & (0.3764)\\

600  (51.83\%) & 0.5298 & 1.0981 & 2.1247  & 0.4521 & $-$1.0290 & 0.6109 &  0.3615\\
&  (0.1554)  & (0.5518) & (0.7656) & (0.1263) & (0.1664)  & (0.4603) & (0.2912)\\

800 (51.72\%) & 0.5203 & 1.0777 & 2.1129 & 0.4698 & $-$1.0224 & 0.6152 & 0.3654 \\
&  (0.1220) & (0.4531) & (0.6628) & (0.1069) & (0.1426)  & (0.4021) & 0.2560\\

1000 (51.85\%) & 0.5186 & 1.0587 & 2.0571   & 0.4708   & $-$1.0083 & 0.6154   & 0.3686\\
&  (0.1103)  & (0.3999) & (0.5986) & (0.0953) & (0.1269)  & (0.3631) & (0.2331)\\
 \hhline{=::=======}
200 (65.85\%) & 1.1006 & 11.5546 & 2.5208 & 0.4488 & $-$1.1013 & 0.6099 & 0.3526\\	
& (0.2953) & (4.1330) & (1.9962) & (0.3050) & (0.3637)	& (1.1864) & (0.7663)\\

400 (64.12\%) & 1.0700 & 10.1602 & 2.3147 & 0.4553 & $-$1.0250 & 0.6108 & 0.3630\\	
& (0.1843) & (2.4199) & (1.3223) & (0.1996) & (0.2449) & (0.7813) & (0.5991)\\

600 (63.65\%) & 1.0697 & 10.0805 & 2.1491 & 0.4570 & $-$1.0177 & 0.6120 & 0.3637\\	
& (0.1393) & (1.9024) & (1.0618) & (0.1566) & (0.1966) & (0.6286) & (0.4134)\\

800 (63.49\%) & 1.0480 & 10.0697 & 2.1451 & 0.4724 & $-$1.0141 & 0.6157 & 0.3638\\	
& (0.1110) & (1.6047) & (0.8821) & (0.1337) & (0.1709) & (0.5263) & (0.3399)\\

1000 (62.98\%) & 1.0260 & 10.0331 & 2.0233 & 0.4853 & $-$1.0059 & 0.6175 & 0.3657\\	
& (0.0907) & (1.0790) & (0.7557) & (0.1163) & (0.1510) & (0.4556) & (0.2930)\\
\multicolumn{1}{c}{\textbf{True Values} $\to$} & $\mathbf{1.000}$&$\mathbf{10.000}$&$\mathbf{2.000}$&$\mathbf{0.500}$&$\mathbf{-1.000}$&$\mathbf{0.623}$&$\mathbf{0.377}$\\
\hline
\end{tabular}
\end{table}

\newpage

\subsection{Application}\label{subsec:app1}
\noindent
The proposed model is now used to analyse a real-world data set corresponding to survival times of $n = 417$ patients; see \cite{llsv:18}. This data set 
presents 56\% of censored observations. The covariates associated with each patient, $i = 1, \ldots, 417$, are:
\begin{itemize}
\item $t_{i}$: time (in years); 
\item $x_{i1}$: treatment with 0 for observation and 1 for interferon);
\item $x_{i2}$: age (in years); 
\item $x_{i3}$: nodule (nodule category: 1 to 4);
\item $x_{i4}$: sex (0 for male and 1 for female);
\item $x_{i5}$: p.s. (performance status-patient's functional capacity scale
as regards his daily activities: 0 for fully active and 1 for other); and
\item $x_{i6}$: tumor (tumor thickness in $mm$). 
\end{itemize}
The interest here lies in the effects of these covariates on the cure fraction. Table~\ref{Table_fitapp1}
presents the estimates of the parameters and their respective standard errors
for the mixture cure Beta-Prime (MBP) and NBBP cure rate regression models. Note that we consider the NBBP regression
model with all covariates, but only the covariate $x_{i3}$,
which represents nodule category, was significant. Since the values of the Akaike information criterion (AIC) and Bayesian information criterion (BIC) are smaller for the NBBP model compared with those values of the MBP model, this new distribution seems to be a very competitive model for these data.


\begin{table}[!ht]
\centering
\caption{ ML estimate (SE in parentheses) and selection
criteria to different models.}\label{Table_fitapp1}
\begin{tabular}{crrcrrrc}
\hline\\[-0.4cm]
 \multirow{2}{*}{Parameter}         &  \multicolumn{3}{c}{MBP} && \multicolumn{3}{c}{NBBP}  \\\cline{2-4} \cline{6-8}\\[-0.4cm]
                   & \multicolumn{1}{c}{MLE}        & \multicolumn{1}{c}{SE}    & $p$-value &&   \multicolumn{1}{c}{MLE}     &  \multicolumn{1}{c}{SE}  & $p$-value\\
\hline
$\alpha$           & \multicolumn{1}{c}{$-$}        & \multicolumn{1}{c}{$-$}   & $-$       && 2.778    & 1.075     & $-$\\
$\mu$              & 3.858      & 0.599 & $-$       && 3.693    & 0.539     & $-$\\
$\phi$             & 2.085      & 0.514 & $-$       && 2.054    & 0.279     & $-$\\
$\beta_{0}$        & 2.159      & 0.802 & 0.007     && $-$0.873 & 2.345     & 0.709\\
$\beta_{1}$        & $-$0.179   & 0.273 & 0.510     && $-$0.076 & 0.115     & 0.505\\
$\beta_{2}$        & $-$0.021   & 0.011 & 0.059     && $-$0.008 & 0.005     & 0.158\\
$\beta_{3}$        & $-$1.382   & 0.488 & 0.004     && $-$0.616 & 0.290     & 0.033\\
$\beta_{4}$        &    0.260   & 0.278 & 0.349     && 0.056    & 0.134     & 0.675\\
$\beta_{5}$        & $-$0.171   & 0.417 & 0.682     && $-$0.114 & 0.166     & 0.489\\
$\beta_{6}$        & $-$0.116   & 0.078 & 0.138     && $-$0.017    & 0.019  & 0.396\\ \hhline{========}
\textbf{AIC}                & \multicolumn{3}{c}{1049.353}   && \multicolumn{3}{c}{1045.886} \\
\textbf{BIC}                & \multicolumn{3}{c}{1086.650}    && \multicolumn{3}{c}{1084.217}\\
\hline
\end{tabular}
\end{table}

\begin{figure}[!ht]
\centering
\psfrag{0}[c][c]{\scriptsize{0.0}}
\psfrag{1}[c][c]{\scriptsize{1.0}}
\psfrag{2}[c][c]{\scriptsize{2.0}}
\psfrag{3}[c][c]{\scriptsize{3.0}}
\psfrag{4}[c][c]{\scriptsize{4.0}}
\psfrag{5}[c][c]{\scriptsize{5.0}}
\psfrag{6}[c][c]{\scriptsize{6.0}}
\psfrag{7}[c][c]{\scriptsize{7.0}}
\psfrag{0.0}[c][c]{\scriptsize{0.0}}
\psfrag{0.2}[c][c]{\scriptsize{0.2}}
\psfrag{0.4}[c][c]{\scriptsize{0.4}}
\psfrag{0.5}[c][c]{\scriptsize{0.5}}
\psfrag{0.6}[c][c]{\scriptsize{0.6}}
\psfrag{0.8}[c][c]{\scriptsize{0.8}}
\psfrag{1.0}[c][c]{\scriptsize{1.0}}
\psfrag{1.5}[c][c]{\scriptsize{1.5}}
\psfrag{-3}[c][c]{\scriptsize{$-$3.0}}
\psfrag{-2}[c][c]{\scriptsize{$-$2.0}}
\psfrag{-1}[c][c]{\scriptsize{$-$1.0}}
\psfrag{fsf}[c][c]{\scriptsize{survival function}}
\psfrag{x}[c][c]{\scriptsize{survival time}}
\psfrag{eq}[c][c]{\scriptsize{empirical quantile}}
\psfrag{tq}[c][c]{\scriptsize{theoretical quantile}}
\subfigure[{\small QQ-NBBP}]{\includegraphics[height=5.5cm,width=4.5cm,angle=-90]{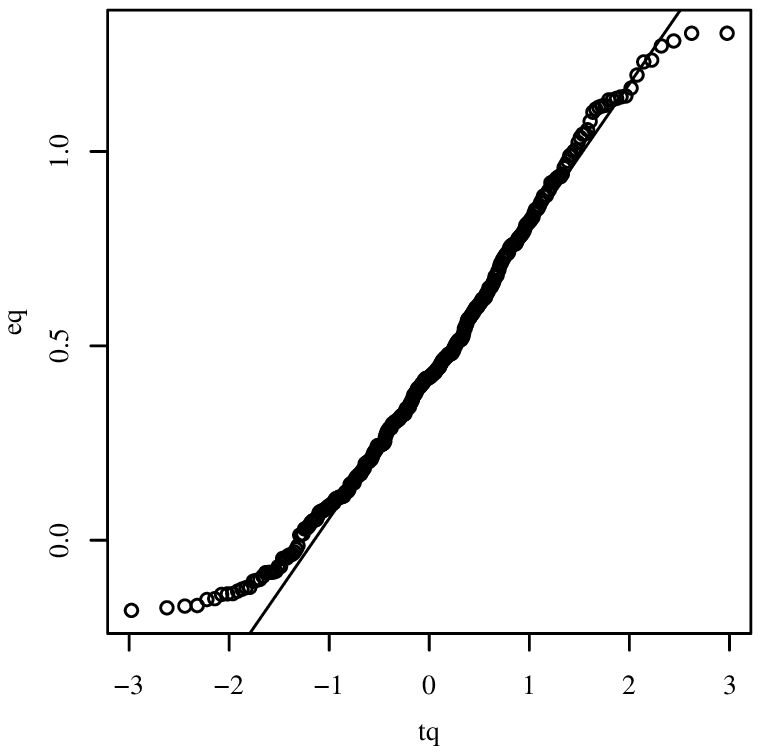}}
\subfigure[{\small SF-NBBP}]{\includegraphics[height=5.5cm,width=4.5cm,angle=-90]{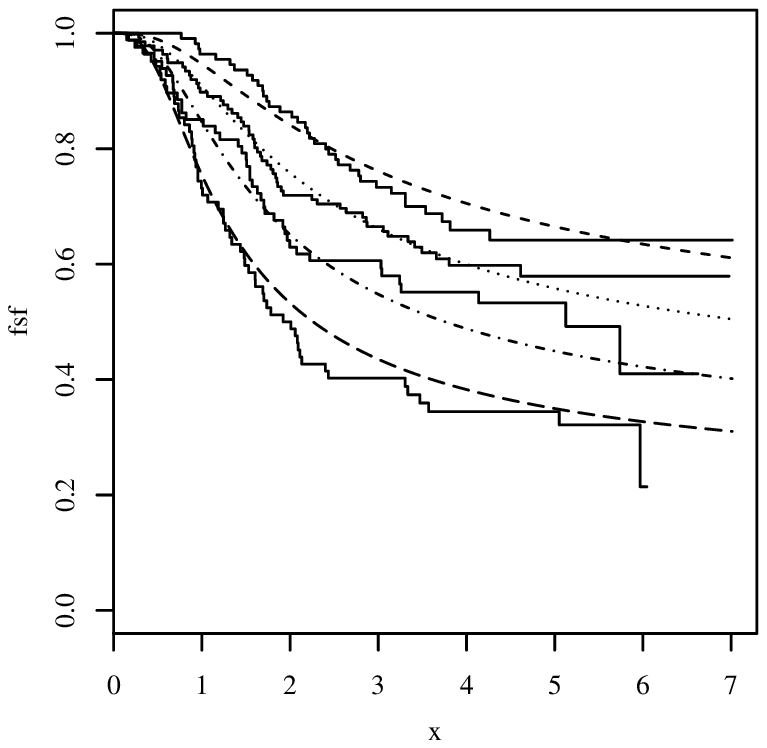}}\\
\subfigure[{\small QQ-MBP}]{\includegraphics[height=5.5cm,width=4.5cm,angle=-90]{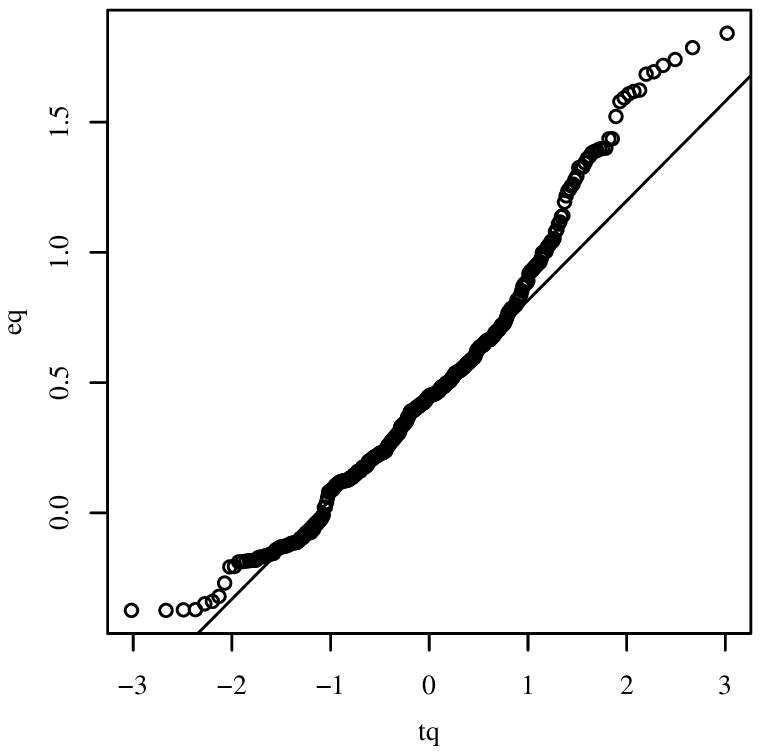}}
\subfigure[{\small SF-MBP}]{\includegraphics[height=5.5cm,width=4.5cm,angle=-90]{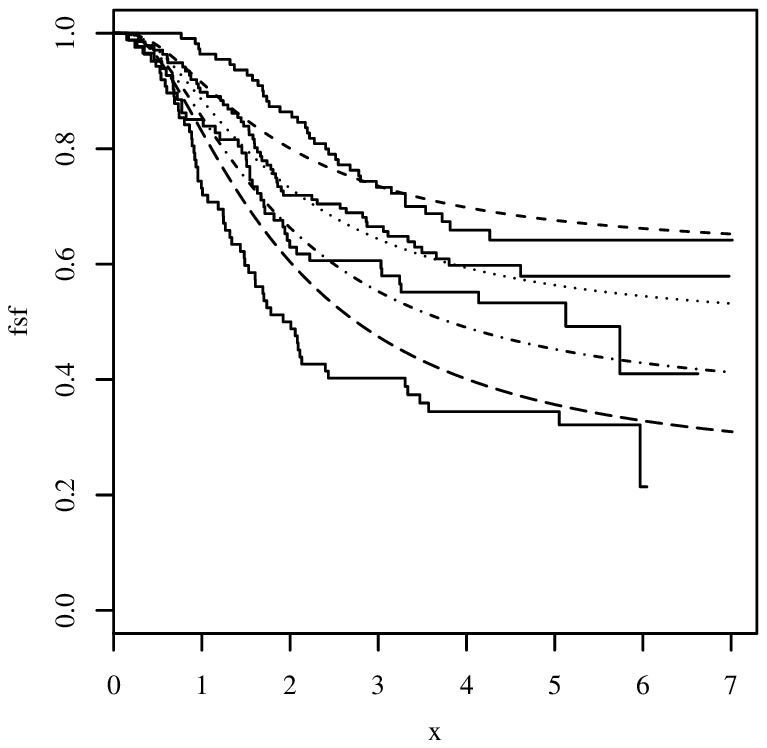}}
\caption{(a) QQ plot (a) and fitted KM (b) for the NBBP model and
QQ plot (c) and fitted KM (d) for the MBP model.}
  \label{fig:app1}
\end{figure}

Figure~\ref{fig:app1} displays the QQ-plots of the normalized randomized quantile
residuals (each point corresponds to the median of five sets of ordered residuals) and the fitted SFs for nodal category based on the Kaplan-Meier (KM) estimator, for the NBBP
and MBP regression models. From this figure, observe that the models are quite similar in terms of the fitted SFs, however the residuals show better agreement in the
NBBP model, which corroborates the results in Table~\ref{Table_fitapp1}.

\begin{figure}[!ht]
\centering
\psfrag{0.0}[c][c]{\scriptsize{0.0}}
\psfrag{0.1}[c][c]{\scriptsize{0.1}}
\psfrag{0.2}[c][c]{\scriptsize{0.2}}
\psfrag{0.3}[c][c]{\scriptsize{0.3}}
\psfrag{0.4}[c][c]{\scriptsize{0.4}}
\psfrag{0.5}[c][c]{\scriptsize{0.5}}
\psfrag{0.6}[c][c]{\scriptsize{0.6}}
\psfrag{0.7}[c][c]{\scriptsize{0.7}}
\psfrag{0.8}[c][c]{\scriptsize{0.8}}
\psfrag{1.0}[c][c]{\scriptsize{1.0}}
\psfrag{0.00}[c][c]{\scriptsize{0.00}}
\psfrag{0.01}[c][c]{\scriptsize{0.01}}
\psfrag{0.02}[c][c]{\scriptsize{0.02}}
\psfrag{0.03}[c][c]{\scriptsize{0.03}}
\psfrag{0.04}[c][c]{\scriptsize{0.04}}
\psfrag{0.05}[c][c]{\scriptsize{0.05}}
\psfrag{0.06}[c][c]{\scriptsize{0.06}}
\psfrag{0.07}[c][c]{\scriptsize{0.07}}
\psfrag{0.08}[c][c]{\scriptsize{0.08}}
\psfrag{0.09}[c][c]{\scriptsize{0.09}}
\psfrag{0.10}[c][c]{\scriptsize{0.10}}
\psfrag{0.11}[c][c]{\scriptsize{0.11}}
\psfrag{0.12}[c][c]{\scriptsize{0.12}}
\psfrag{0.14}[c][c]{\scriptsize{0.14}}
\psfrag{0.15}[c][c]{\scriptsize{0.15}}
\psfrag{0.20}[c][c]{\scriptsize{0.20}}
\psfrag{0}[c][c]{\scriptsize{0}}
\psfrag{1}[c][c]{\scriptsize{1}}
\psfrag{2}[c][c]{\scriptsize{2}}
\psfrag{3}[c][c]{\scriptsize{3}}
\psfrag{4}[c][c]{\scriptsize{4}}
\psfrag{5}[c][c]{\scriptsize{5}}
\psfrag{6}[c][c]{\scriptsize{6}}
\psfrag{8}[c][c]{\scriptsize{8}}
\psfrag{10}[c][c]{\scriptsize{10}}
\psfrag{15}[c][c]{\scriptsize{15}}
\psfrag{20}[c][c]{\scriptsize{20}}
\psfrag{25}[c][c]{\scriptsize{25}}
\psfrag{30}[c][c]{\scriptsize{30}}
\psfrag{40}[c][c]{\scriptsize{40}}
\psfrag{50}[c][c]{\scriptsize{50}}
\psfrag{60}[c][c]{\scriptsize{60}}
\psfrag{80}[c][c]{\scriptsize{80}}

\psfrag{100}[c][c]{\scriptsize{100}}
\psfrag{120}[c][c]{\scriptsize{120}}
\psfrag{140}[c][c]{\scriptsize{140}}
\psfrag{150}[c][c]{\scriptsize{150}}
\psfrag{200}[c][c]{\scriptsize{200}}
\psfrag{300}[c][c]{\scriptsize{300}}
\psfrag{400}[c][c]{\scriptsize{400}}
\psfrag{500}[c][c]{\scriptsize{500}}
\psfrag{600}[c][c]{\scriptsize{600}}
\psfrag{a}[r][c]{\scriptsize{\#88\,\,\,}}
\psfrag{b}[r][c]{\scriptsize{\#174\,\,\,\,}}
\psfrag{c}[r][c]{\scriptsize{\#247\,\,\,}}
\psfrag{d}[r][c]{\scriptsize{\#279\,\,\,}}
\psfrag{id}[c][c]{\scriptsize{index}}
\psfrag{CxA}[c][c]{\scriptsize{$C_{i}(\alpha)$}}
\psfrag{CxX}[c][c]{\scriptsize{$C_{i}(\bm{\xi})$}}
\psfrag{CxB}[c][c]{\scriptsize{$C_{i}(\bm{\beta})$}}
{\includegraphics[height=5.0cm,width=5.5cm,angle=-90]{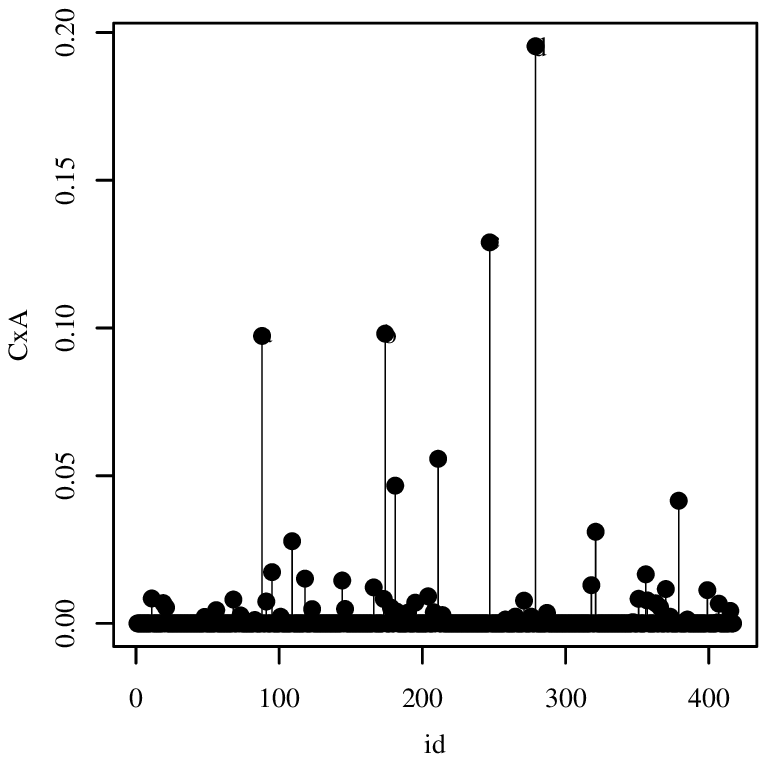}}
{\includegraphics[height=5.0cm,width=5.5cm,angle=-90]{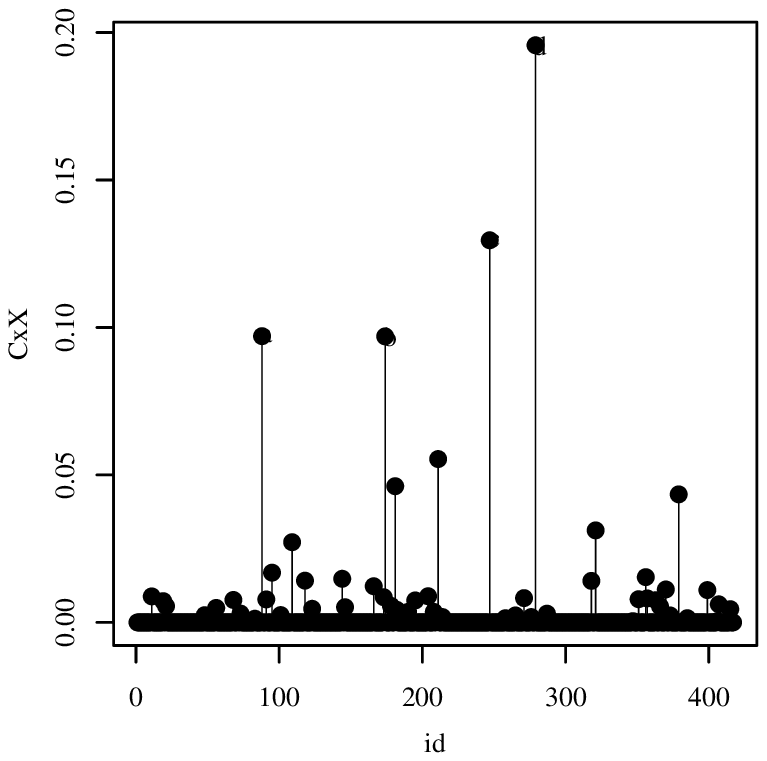}}
{\includegraphics[height=5.0cm,width=5.5cm,angle=-90]{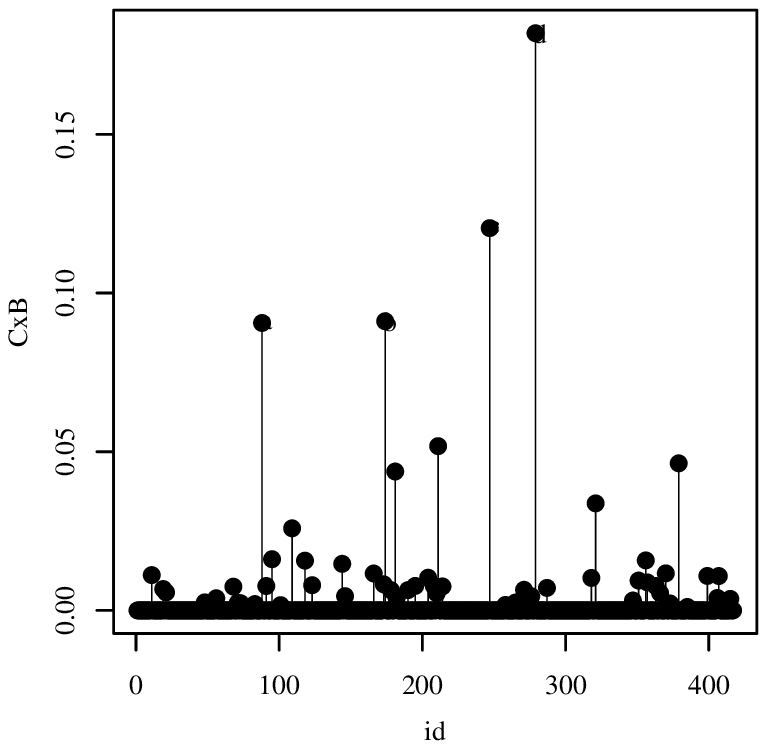}}
\caption{{Index plots of $C_{i}$ for $\alpha$ (left), $\bm{\xi} = (\mu, \phi)^{\top}$ (center)
and $\bm{\beta}$ (right) with case-weight perturbation and melanoma data.}}
 \label{fig:02}
\end{figure}

Index plots of $C_{i}$ under case-weight perturbation are shown in Figure~\ref{fig:02}. We omit the plots corresponding to response and covariate (tumor thickness, $x_{i6}$) perturbations as they look very similar. Note that the cases \#88, \#174, \#247 and \#279 are detected as potential influential observations under the considered perturbation schemes. 

\newpage 

We assess the
impact of the detected influential cases on the model inference with the relative change (RC), which is obtained by removing influential cases and re-estimating the parameters and their corresponding SEs.
The RCs in the parameter estimates and their corresponding estimated SEs are shown in Table~\ref{tab:5}. Also, $p$-values from the associated $t$-test are shown for the regression coefficients. From
this table, note that, in general, the largest RCs are related to the removal of the case \#279. Moreover, inferential changes are found only on $\widehat{\beta}_{3}$.

\begin{table}[!ht]
\centering
\small
\renewcommand{\arraystretch}{0.55}
\renewcommand{\tabcolsep}{0.01cm}
\caption{RC (in \%) in ML estimates and their corresponding SEs, and respective $p$-values in brackets; inferential changes are highlighted in gray.}\label{tab:5}
 \begin{tabular}{clcccccccccc}
\toprule
 Dropped case(s)&  & $\widehat{\mu}$ & $\widehat{\phi}$  & $\widehat{\alpha}$ & $\widehat{\beta}_{0}$ & $\widehat{\beta}_{1}$ & $\widehat{\beta}_{2}$ & $\widehat{\beta}_{3}$ & $\widehat{\beta}_{4}$ & $\widehat{\beta}_{5}$ & $\widehat{\beta}_{6}$\\
\midrule
\\[-0.2cm]
\{\#88\} & $\textrm{RC}_{\zeta_{j(i)}}$  & 8.43 & 96.25  &  291.90 & 239.44 &  828.98  & 2.75  &  25.82 & 76.24 & 52.64  &  49.07\\
& $\textrm{RC}_{\textrm{SE}(\zeta_{j(i)})}$  & (22.5) & (89.45) & (68.62)  & (70.95) & (63.17) & (31.01) & (17.62)  & (51.32)  & (45.19)  & (65.20)\\
& $p$-value  & - & -  & - &  [0.651] & [0.567] & [0.112]  & \cellcolor[gray]{0.9}[0.014] & [0.692]  & [0.644] & [0.380]\\

\{\#174\} & $\textrm{RC}_{\zeta_{j(i)}}$  & 6.66 & 94.60 & 294.73 & 210.75 & 765.32 & 2.31 & 23.34 & 62.06 & 57.02 &  46.90\\
& $\textrm{RC}_{\textrm{SE}(\zeta_{j(i)})}$  & (15.29) & (63.43) & (53.87) & (318.64) & (62.41) & (22.69) & (87.72) & (31.16) & (42.17) & (58.56)\\
& $p$-value  & - & -  & - & [0.883] & [0.601] & [0.347] & \cellcolor[gray]{0.9}[0.264] & [0.654] & [0.690] & [0.442]\\

\{\#247\} & $\textrm{RC}_{\zeta_{j(i)}}$  & 8.27 & 96.79 & 296.26 & 254.46 & 851.30 & 10.69 & 27.28 & 77.23 & 51.78 & 51.58\\
& $\textrm{RC}_{\textrm{SE}(\zeta_{j(i)})}$  & (23.07) & (94.10) & (69.8) & (11.08) & (63.76) & (38.67) & (33.98) & (53.16) & (46.06) & (66.21)\\
& $p$-value  & - & -  & - & [0.440] & [0.551] & [0.101]  & [0.003] & [0.693] & [0.632] & [0.390]\\

\{\#279\} & $\textrm{RC}_{\zeta_{j(i)}}$  & 2.00 & 96.60 & 344.63 & 241.57 & 974.54 & 21.37 & 29.49 & 84.83 & 64.33 & 53.94\\
& $\textrm{RC}_{\textrm{SE}(\zeta_{j(i)})}$  & (14.49) & (91.94) & (66.35) & (30.25) & (65.03) & (40.43) & (30.21) & (54.36) & (48.88) & (67.54)\\
& $p$-value  & - & -  & - & [0.546] & [0.485] & [0.137] & [0.005] & [0.787] & [0.708] & [0.394]\\

\{\#88,\#174\} & $\textrm{RC}_{\zeta_{j(i)}}$  & 6.74 & 94.47 & 289.13 & 207.90  & 638.35 & 7.46 & 23.58  & 64.44 & 82.37 & 46.37\\
& $\textrm{RC}_{\textrm{SE}(\zeta_{j(i)})}$  & (17.64) & (70.43) & (59.41) & (235.80) & (62.28) & (8.05) & (52.74) & (38.93) & (43.81) & (60.66)\\
& $p$-value  & - & -  & - & [0.858] & [0.656] & [0.263] & \cellcolor[gray]{0.9}[0.171] & [0.636] & [0.866] & [0.413]\\

\{\#88,\#247\} & $\textrm{RC}_{\zeta_{j(i)}}$  & 8.50 & 95.93 & 286.03 & 237.18 & 715.64 & 3.52 & 26.04 & 78.82 & 76.54 & 49.76\\
& $\textrm{RC}_{\textrm{SE}(\zeta_{j(i)})}$  & (22.91) & (89.24) & (68.60) & (62.84) & (62.78) & (30.47) & (17.90) & (51.16) & (44.61) & (64.84)\\
& $p$-value  & - & -  & - & [0.639] & [0.618] & [0.118]  & [0.013] & [0.724] & [0.820] & [0.391]\\

\{\#88,\#279\} & $\textrm{RC}_{\zeta_{j(i)}}$  & 2.26 & 95.61 & 334.19 & 225.00 & 872.31 & 15.32 & 28.42 & 86.32 & 87.94 & 52.77\\
& $\textrm{RC}_{\textrm{SE}(\zeta_{j(i)})}$  & (15.41) & (87.55) & (65.23) & (60.46) & (64.20) & (33.73) & (16.65) & (52.61) & (47.48) & (66.35)\\
& $p$-value  & - & -  & - & [0.665] & [0.537] & [0.150] & [0.018] & [0.814] & [0.902] & [0.399]\\

\{\#174,\#247\} & $\textrm{RC}_{\zeta_{j(i)}}$  & 6.25 & 90.13 & 277.20 & 169.98 & 658.18 & 6.10 & 19.87 & 60.76 & 80.13 & 45.66\\
& $\textrm{RC}_{\textrm{SE}(\zeta_{j(i)})}$  & (3.08) & (24.97) & (17.79) & (413.55) & (59.30) & (88.09) & (190.72) & (1.31) & (38.08) & (41.98)\\
& $p$-value  & - & -  & - & [0.939] & [0.672] & [0.525] & \cellcolor[gray]{0.9}[0.451] & [0.746] & [0.863] & [0.574]\\

\{\#174,\#279\} & $\textrm{RC}_{\zeta_{j(i)}}$  & 6.84 & 11.90 & 36.39 & 92.60 & 1490.16 & 75.48 & 46.58 & 10.56 & 73.46 & 84.34\\
& $\textrm{RC}_{\textrm{SE}(\zeta_{j(i)})}$  & (4.92) & (22.66) & (29.10) & (44.67) & (2.80) & (28.98) & (79.37) & (10.18) & (9.86) & (41.56)\\
& $p$-value  & - & -  & - & [0.459] & [0.758] & [0.125] & [0.025] & [0.418] & [0.897] & [0.434]\\

\{\#247,\#279\} & $\textrm{RC}_{\zeta_{j(i)}}$  & 10.52 & 28.55 & 3.76 & 56.10 & 1261.59 & 75.14 & 33.02 & 21.34 & 86.37 & 52.08\\
& $\textrm{RC}_{\textrm{SE}(\zeta_{j(i)})}$  & (11.44) & (34.82) & (36.27) & (48.27) & (3.90) & (43.15) & (69.34) & (9.97) & (4.45) & (16.19)\\
& $p$-value  & - & -  & - & [0.558] & [0.783] & [0.168]  & [0.031] & [0.478] & [0.944] & [0.432]\\

\{\#88,\#174,\#247\} & $\textrm{RC}_{\zeta_{j(i)}}$  & 6.71 & 95.29 & 288.71 & 223.04 & 528.73 & 3.58 & 25.44 & 68.78 & 107.75 & 48.64\\
& $\textrm{RC}_{\textrm{SE}(\zeta_{j(i)})}$  & (22.09) & (88.18) & (68.81) & (56.75) & (62.65) & (29.07) & (17.01) & (50.43) & (44.40) & (64.81)\\
& $p$-value  & - & -  & - & [0.662] & [0.702] & [0.100]  & [0.014] & [0.609] & [0.940] & [0.381]\\

\{\#88,\#174,\#279\} & $\textrm{RC}_{\zeta_{j(i)}}$ & 0.19 & 93.91 & 334.44 & 198.78 & 689.95 & 5.89 & 26.61 & 75.03 & 119.53 & 50.44\\
& $\textrm{RC}_{\textrm{SE}(\zeta_{j(i)})}$  & (12.92) & (76.44) & (59.99) & (125.73) & (63.56) & (16.19) & (17.10) & (47.06) & (46.21) & (64.07)\\
& $p$-value  & - & -  & - & [0.808] & [0.622] & [0.206] & \cellcolor[gray]{0.9}[0.086] & [0.701] & [0.845] & [0.408]\\

\{\#174,\#247,\#279\} & $\textrm{RC}_{\zeta_{j(i)}}$  & 18.47 & 46.79 & 15.35 & 59.59 & 2261.28 & 93.71 & 49.00  & 36.77 & 19.45 & 80.57\\
& $\textrm{RC}_{\textrm{SE}(\zeta_{j(i)})}$  & (33.76) & (69.75) & (60.28) & (18.49) & (7.41) & (40.15) & (81.51) & (13.25) & (25.73) & (10.75)\\
& $p$-value  & - & -  & - & [0.454] & [0.596] & [0.119]  & [0.024] & [0.554] & [0.610] & [0.328]\\

\{\#88,\#174,\#247,\#279\} & $\textrm{RC}_{\zeta_{j(i)}}$  & 21.12 & 49.09 & 34.49 & 46.31 & 2013.96 & 86.00 & 52.86 & 19.98 & 59.35 & 127.67\\
& $\textrm{RC}_{\textrm{SE}(\zeta_{j(i)})}$  & (30.85) & (13.59) & (8.92) & (230.86) & (54.93) & (113.29) & (368.31) & (10.23) & (43.73) & (245.70)\\
& $p$-value  & - & -  & - & [0.806] & [0.779] & [0.326]  & \cellcolor[gray]{0.9}[0.372] & [0.555] & [0.879] & [0.692]\\

\bottomrule
\end{tabular}
\end{table}

\section{Concluding remarks}\label{sec_conc}
\noindent

In this paper, we proposed a new model for survival data assuming competing causes of the event
of interest follows the negative binomial distribution and the time to event
follow a BP distribution. Estimation was approached by the maximum likelihood method. Diagnostic tools have been obtained to detect locally influential
observations in the maximum likelihood estimates. A Monte Carlo simulation study was carried out to evaluate the behavior of the proposed model
parameters. In the application
to a medical real-world data set, we observed that the new cure rate model delivers the best fit. We hope the
proposed model attracts the attention of practitioners of survival analysis.




\end{document}